\newcommand{\beq}{\begin{equation}}
\newcommand{\eeq}{\end{equation}}
\newcommand{\bea}{\begin{eqnarray}}
\newcommand{\eea}{\end{eqnarray}}

\newcommand{\gsim}{\lower.7ex\hbox{$\;\stackrel{\textstyle>}{\sim}\;$}}
\newcommand{\lsim}{\lower.7ex\hbox{$\;\stackrel{\textstyle<}{\sim}\;$}}

\documentclass[aps,prev,twocolumn,preprintnumbers,floatfix,nofootinbib]{revtex4-1}
\usepackage{graphicx}
\usepackage{epstopdf}
\usepackage{mathrsfs}
\usepackage{amssymb}
\usepackage{verbatim}

\def\stacksymbols #1#2#3#4{\def\theguybelow{#2}
    \def\vp{\lower#3pt}
    \def\sp{\baselineskip0pt\lineskip#4pt}
    \mathrel{\mathpalette\intermediary#1}}

\def\intermediary#1#2{\vp\vbox{\sp
     \everycr={}\tabskip0pt
     \halign{$\mathsurround0pt#1\hfil##\hfil$\crcr#2\crcr
              \theguybelow\crcr}}}

\def\be{\begin{equation}}
\def\ee{\end{equation}}
\def\bea{\begin{eqnarray}}
\def\eea{\end{eqnarray}}

\def\sp{\;\;\;,\;\;\;}

\def\lsim{\raise0.3ex\hbox{$\;<$\kern-0.75em\raise-1.1ex\hbox{$\sim\;$}}}
\def\gsim{\raise0.3ex\hbox{$\;>$\kern-0.75em\raise-1.1ex\hbox{$\sim\;$}}}

\def\inbar{\,\vrule height1.5ex width.4pt depth0pt}

\def\IC{\relax\hbox{$\inbar\kern-.3em{\rm C}$}}
\def\IQ{\relax\hbox{$\inbar\kern-.3em{\rm Q}$}}
\def\IR{\relax{\rm I\kern-.18em R}}
 \font\cmss=cmss10 \font\cmsss=cmss10 at 7pt
\def\IZ{\relax\ifmmode\mathchoice
 {\hbox{\cmss Z\kern-.4em Z}}{\hbox{\cmss Z\kern-.4em Z}}
 {\lower.9pt\hbox{\cmsss Z\kern-.4em Z}}
 {\lower1.2pt\hbox{\cmsss Z\kern-.4em Z}}\else{\cmss Z\kern-.4em Z}\fi}

\def\comment#1{}

\def\u1x{U(1)_X}
\newcommand{\nc}{\newcommand}
\nc{\LL}{L}
\nc{\vv}{\tilde{v}}
\nc{\ccdot}{\!\cdot\!}
\nc{\gsm}{G_{SM}}
\nc{\vfive}{\mathbf{5}\oplus\mathbf{\overline{5}}}
\nc{\vten}{\mathbf{10}\oplus\mathbf{\overline{10}}}
\nc{\zhol}{Z^{\rm hol}}
\nc{\xfb}{\,{\rm fb}}

\begin{document}

%
%

\preprint{CERN-PH-TH-2011-271}
\preprint{DESY 11-202}
\preprint{LPT--Orsay 11/90}

\title{Vector Higgs--portal dark matter  and the invisible Higgs}

\author{Oleg Lebedev$^{a}$}
\email{lebedev@mail.desy.de}
\author{Hyun Min Lee$^{b}$}
\email{Hyun.Min.Lee@cern.ch}
\author{Yann Mambrini$^{c}$}
\email{Yann.Mambrini@th.u-psud.fr}

\vspace{0.2cm}
\affiliation{
${}^a$  DESY Theory Group,  Notkestrasse 85, D-22607 Hamburg, Germany }
\affiliation{ 
${}^b$  CERN, Theory Division, CH-1211 Geneva 23, Switzerland   }
\affiliation{
${}^c$ Laboratoire de Physique Th\'eorique 
Universit\'e Paris-Sud, F-91405 Orsay, France}

\begin{abstract}
The Higgs sector of the Standard Model offers a unique probe 
of the hidden sector. In this work, 
we explore the possibility of renormalizable Higgs couplings to the hidden sector
vector fields which can constitute dark matter (DM). Abelian gauge sectors with
minimal field content,  necessary to render the gauge fields massive, have a
natural $Z_2$ parity. This symmetry ensures stability of the vector fields making
them viable dark matter candidates, while evading the usual electroweak constraints.  
We illustrate this idea with the St\"uckelberg and Higgs 
mechanisms. Vector DM is consistent with  the WMAP, XENON100, and LHC constraints, while
it can affect significantly the invisible Higgs decay. Due to the enhanced  branching ratio
for the Higgs decay into the longitudinal components of the vector field, the vector Higgs
portal provides an efficient way to hide the Higgs at the LHC. This could be the reason
why  the latest combined  ATLAS/CMS  data  did not bring  
evidence for  the existence of the Higgs boson.
\end{abstract}

\maketitle


\maketitle


\setcounter{equation}{0}



\section{Introduction}

Two of the most important issues in particle physics phenomenology are the nature
 of  dark matter and understanding  electroweak symmetry breaking.
While  85 \% of matter is dark \cite{WMAP}, it remains elusive to direct detection 
and the  Weakly Interacting Massive Particle (WIMP) paradigm becomes more and more 
constrained  \cite{Aprile:2011ts}.
On the other front, the accelerator collaborations 
ATLAS \cite{ATLAS,COMBI}, CMS \cite{CMS} and D0/CDF \cite{D0, TEVATRON}
have obtained important results concerning the Higgs searches.
If the Higgs boson is the main communicator between the dark world and ours,
Higgs hunting  is intimately related  to direct DM detection.
To reconcile all the constraints is then a  non--trivial task \cite{Mambrini:2011ik}
(see also \cite{Mambrini:2011pw} for more general scenarios).
Whereas this idea is normally considered in the context of scalar (or fermion) dark matter,
 in this work, we study the possibility
of vectorial  dark matter  interacting with  the visible sector  through the Higgs portal.  
We show that this framework is well motivated theoretically and, while
being consistent with the existing constraints, can affect 
significantly the ongoing Higgs search at the LHC.

\noindent
The paper is organized as follows:
in section II we introduce  the vector Higgs portal
and provide its theoretical basis; in section III we study relevant 
dark matter and accelerator constraints, and discuss implications
for the LHC Higgs search; our conclusions are presented in section IV.

\section{Vector Higgs portal }

The Higgs sector offers a unique opportunity to probe the hidden sector.
The operator $\bar H H$ is the only dim-2 gauge and Lorentz 
invariant operator in the Standard model.
Therefore renormalizable interactions of the form 
\begin{equation}
\Delta {\cal L}_{\rm scalar}= {\lambda_{hs} \over 4} ~\bar H H ~( S^2 + m_{hs} S )
\label{scalar-portal}
\end{equation}
are possible, where $S$ is a hidden sector (real) scalar. 
These are known as the scalar  ``Higgs portal'' (the name was coined in \cite{Patt:2006fw}).
More generally, one can also have a dim-4   vector Higgs portal, which couples
the Higgs doublet to a massive    
vector field $X_\mu$ from the hidden sector,
\begin{eqnarray}
\Delta {\cal L}_{\rm vector} &=& {\lambda_{hv} \over 4} ~ \bar H H~ X_\mu X^\mu + ( \xi_1 ~
\bar H iD_\mu H ~ X^\mu  \nonumber\\
&+& \xi_2 ~ \bar H H~ i \partial_\mu X^\mu +
{\rm h.c.}) \;, 
\label{vector-portal}
\end{eqnarray}
where the covariant derivative $D_\mu$  is taken with respect to the Standard Model (SM) gauge
group.\footnote{For generality, we have included the $\partial_\mu X^\mu$ term,
which may be relevant in models with field--dependent mass terms. }
 $X_\mu$ can be associated with a hidden U(1). It  becomes massive
due to the Higgs or St\"uckelberg mechanism in the hidden sector. 

A hidden vector field can be a good dark matter candidate. 
Suppose its decay into the hidden sector is kinematically forbidden 
and the only communication with the SM is given by Eq.~\ref{vector-portal}.
It is then stable
if $\xi_{1,2}$ are zero, 
in which case the Higgs portal has a $Z_2$ symmetry
$X_\mu \rightarrow - X_\mu$. In principle, $X_\mu$  can also be made
long--lived by adjusting $\xi_1$ and $\xi_2$ to be very small. 
In particular, if 
there are  fields charged under the U(1) apart from those providing longitudinal
components of the gauge bosons, these  terms are  usually generated and 
it requires non--trivial engineering to make $X_\mu$ stable
on the cosmological scale. In the absence of such charged fields,  
 the Lagrangian enjoys a natural
 $Z_2$ parity which makes $X_\mu$ stable. 
This applies both to the Higgs and the  St\"uckelberg mechanisms.
Let us consider   examples, starting with the less common  St\"uckelberg dark matter.

\subsection{ Examples: St\"uckelberg  and ``Higgsed'' dark matter }

Consider a U(1) vector field which becomes massive through the 
 St\"uckelberg mechanism (for a review, see \cite{Ruegg:2003ps}),
\begin{equation}
{\cal L}_{\rm St} = -{1\over 4} F_{\mu\nu} F^{\mu\nu} + {1\over 2} m^2
X_\mu X^\mu \;,
\end{equation}
where $X_\mu$ is composed of a vector potential $X_\mu'$ and a
 St\"uckelberg axion--like field $\phi$ with the following gauge
transformation rule
\begin{eqnarray}
&& X_\nu \equiv  X_{\nu}' + {1\over \mu} \;  \partial_\nu \phi \;, \nonumber\\
&& \delta X_\nu' = \partial_\nu \epsilon  \;, \nonumber\\
&& \delta \phi = - \mu \epsilon \;.
\end{eqnarray}
Here $\mu$ is a mass scale and $\epsilon $ is a gauge transfromation 
parameter. Note that the Lagrangian possesses a symmetry $X_\mu \rightarrow - X_\mu$,
which in terms of the original variables reads 
 $X_\mu' \rightarrow - X_\mu'$, $\phi \rightarrow - \phi$.\footnote{This symmetry is
also preserved by the gauge fixing term.}
Taking $\mu=m$ corresponds to choosing  a canonical kinetic term
for the axion. 
In general, however,  $m^2$ can be field--dependent, in which case 
the axion kinetic term involves a function of other fields of the system.
Expressing
\begin{equation}  
m^2 = \mu^2 f(S) \;,
\end{equation}
with $S$ being a  scalar, we  recover the axion 
kinetic term $1/2 \; f(S) (\partial_\mu \phi)^2 $ in the massless
limit $\mu \rightarrow 0$. Consider 2 simple possibilities.

{\it  \bf (i) }  One may  allow for  a Higgs--dependent mass term 
\begin{equation}
m^2 = \mu^2 \left( 1  +  \zeta ~ {\bar H H \over M^2} \right) \;, 
\end{equation}
where $M$ is another scale. 
In this case, the Higgs portal coupling
$\bar H H X_\mu X^\mu$ is generated with
\begin{equation} 
\lambda_{hv} =  2\zeta ~ { \mu^2 \over  M^2 } \;.
\end{equation} 
{\it A priori} neither of these scales is related to the EW scale, nor
to each other so that $\lambda_{hv}$ can be significant. 

{\it \bf (ii)}
Consider another example, where  $\bar H H X_\mu X^\mu$ is generated
through the singlet  Higgs portal.  Take 
\begin{equation}
m^2 = \mu^2 ~ { S^2 \over M^2 }   \;,
\end{equation}
where $S$ is a hidden sector scalar with a potential $V(S)$
and a Higgs portal coupling  $(\lambda_{hs}/4) \bar H H S^2 $.
We have factored out $\mu^2$ in order to have a smooth massless limit.
Obviously, integrating out heavy $S$ at tree level produces the 
required coupling with 
\begin{equation}
\lambda_{hv} = 2\lambda_{hs}~ { \mu^2 \langle S \rangle^2 \over  M^2 m_S^2 } \;,
\end{equation}
where $m_S$ is the mass of $S$. Note that $\lambda_{hv}$ can also be written 
as $ 2\lambda_{hs} m_X^2 / m_S^2$. The procedure of integrating $S$  out 
is justified as long as  $m_h^2 , m_X^2  \ll m_S^2$. In practise,
these quantities can differ by, say, an order of magnitude so that  
$\lambda_{hv} $ can be as large as ${\cal O}(10^{-1}-1)$. 
In the same fashion,   self--interaction 
\begin{equation}
 {1\over 2} ~{\mu^4 \langle S \rangle^2   \over M^4 m_S^2}~ (X_\mu X^\mu)^2 
\end{equation}
is also induced.
Note also that 
the physical Higgs boson mixes with $S$ in this example, with   the mixing
angle being  of order $v/ \langle S \rangle $.

Furthermore, at one loop, both $\bar H H~ X_\mu X^\mu $ and $ (X_\mu X^\mu)^2 $ are generated with (formally) log--divergent coefficients. 
Keeping in 
mind that we are dealing with an effective theory, these are regularized 
by the cutoff which is expected to  be of order $M$. 
This implies  that $\lambda_{hv}$ can be taken as  a free parameter
at low energies.

As the Higgs doublet develops a VEV, an additional mass term 
$ \lambda_{hv} v^2/2  $ for the vector is generated. It can be of either sign, as seen from the
above examples. Therefore  some cancellation between the original
 St\"uckelberg mass and the Higgs--induced one is  allowed.

It is important that the linear in $X_\mu$ terms vanish in these examples,
\begin{equation}
\xi_1=\xi_2=0 \;.
\end{equation}
This is because, in the absence of fields charged under U(1) 
(apart from the axion absorbed in $X_\mu$),
 the Lagrangian possesses a natural    parity
\begin{equation}
X_\mu \rightarrow - X_\mu \;.
\end{equation}
Note that here we consider a U(1) ``orthogonal'' to the SM gauge group, so there is
no mixing with hypercharge at tree level, nor is it  induced radiatively. Thus,
the new vector boson evades the usual electroweak constraints.

Some of the ingredients of this construction are well known in string theory.
In particular, most of  realistic models (see e.g. \cite{Lebedev:2008un})
involve an ``anomalous'' U(1) whose anomaly is cancelled by the Green--Schwarz
mechanism. In this process, the U(1) gauge boson becomes massive through
the  St\"uckelberg mechanism. The corresponding axion has a dilaton--dependent
kinetic term, so that the vector mass term  in Planck units   is
(see e.g. \cite{Lalak:1999bk})
\begin{equation}
{1\over 4 s^2 } ~ \left( {1\over 2} \delta ~X_\mu - \partial_\mu a      \right)^2 \;,
\end{equation}
where $a$ and $s$ are the axion and the dilaton, respectively, and
$\delta $ is a Green--Schwarz parameter. 
It is therefore quite natural to expect  a field--dependent vector boson mass,
although in typical string constructions this mass is close to the Planck scale.

{\it \bf (iii)}
Finally, one can equally well  use the Higgs mechanism in the hidden sector. 
Suppose we have a U(1) gauge field and a charged complex scalar $S$, which has a scalar
Higgs portal coupling, 
\begin{equation}
{\cal L}_{\rm Higgs}=  -{1\over 4} F_{\mu\nu} F^{\mu\nu} + D_\mu S^* D^\mu S - V(S) +
{\lambda_{hs}\over 4 } \bar H H S^* S \;.
\end{equation}
This system enjoys a symmetry (charge conjugation)
\begin{eqnarray}
 X_\mu' &\rightarrow& - X_\mu' \;, \nonumber\\
 S &\rightarrow&  S^* \;, 
\end{eqnarray}
where $X_\mu'$ is the massless gauge field.\footnote{Note that this 
parity does not generalize to the non--abelian case due to the triple gauge boson vertex.} 
This symmetry is retained in the broken phase,
that is, the massive gauge field can be reversed:  $X_\mu \rightarrow - X_\mu$.
The physical Higgs--like excitation has no  interactions which are odd under $X_\mu \rightarrow - X_\mu$,
so integrating it out in the  unitary gauge will only produce even interactions
and $\xi_{1,2}=0$. On the other hand, $\bar H H X_\mu X^\mu$ is generated at both tree and loop level,
the latter being log--divergent.
 Note that the vector field remains light as long as $g^2 \ll \lambda_S$,
where $g$ is the gauge coupling and $\lambda_S$ is the singlet self--interaction. 
Otherwise, the   analysis is very similar to that of 
the St\"uckelberg case.

It is important to remember that these examples are valid within 
effective field theory. Indeed, we either integrate out a heavy 
``hidden Higgs'' or use higher dimensional operators for the field--dependent 
kinetic terms. The range of validity of the effective description can
be estimated from unitarity considerations. The vector portal coupling 
together with the $m^2$  mass term from the hidden sector 
results in 
\begin{equation}
\Delta {\cal L}_X=\frac{1}{2}m^2_X \Big(1+a\,\frac{2h}{v}+b\,\frac{h^2}{v^2}\Big) X_\mu X^\mu \;,
\label{hXX}
\end{equation} 
where the Higgs field is decomposed as $H^0=v+h/\sqrt{2}$; 
$m^2_X=m^2+\frac{1}{2}\lambda_{hv}v^2$ and 
$a/\sqrt{2}=b=\frac{\lambda_{hv}v^2}{ 4 m^2_X}$. 
This form is useful for comparison with the Standard Model Higgs--vector scattering.
The tree level amplitude for
the process $hh \rightarrow XX$ (and $hX  \rightarrow hX $)
grows with energy as ${\cal A}\sim (a^2 - b) {E^2\over  v^2 }$
 for $E \gg m_X$. Note that in the Standard Model an analogous amplitude 
vanishes, whereas here it is proportional to the ``non--Higgs'' mass term $m^2$.  Requiring 
the partial wave amplitude $a_0$ to be less than 1/2 (see e.g. \cite{Djouadi:2005gi}) and ignoring order 1 factors, we find that the unitarity cutoff is of order 
\begin{equation} 
E \sim \sqrt{16 \pi \over \lambda_{hv}} {m_X^2 \over m} \;. \label{uni}
\end{equation}
Above this energy, additional states should be taken into account.
Further, the amplitude for $XX \rightarrow XX$ proceeding through the Higgs exchange 
approaches a constant at  $E\gg m_h, m_X$, which imposes $m_X > \lambda_{hv} v /\sqrt{8\pi}$. 
Finally, in the presence of $X$ self--interaction, unitarity in $XX \rightarrow XX$ 
requires roughly $E< (8\pi/{\lambda})^{1/4}   m_X$, where $\lambda $ is the $X$ quartic coupling.

We see that there is a fundamental difference between the vector and scalar
Higgs portals: the former is UV--incomplete, whereas the latter is 
UV--complete. In the vector case, some amplitudes grow indefinitely with 
energy meaning that the Higgs portal description breaks down. This 
can be traced back to the derivative couplings of the Goldstone bosons
(or longitudinal polarization vectors growing with energy) and the fact that 
$m^2$ appears as a ``hard'' mass term. 
Only in the full theory, which contains extra states, 
is the ``soft'' origin of $m^2$ manifest. In contrast, the scalar mass term
is not problematic and the scalar Higgs portal is UV--complete.

\section{Phenomenology}

The relevant $Z_2$--symmetric   Lagrangian is given by  
\begin{equation}
{\cal L} =  {\lambda_{hv} \over 4} ~ \bar H H~ X_\mu X^\mu + {1\over 2}m^2 X_\mu X^\mu +
{\lambda \over 4}    (X_\mu X^\mu)^2 \;.
\end{equation}
Note that upon EW symmetry breaking  the mass of the vector field
is $m_X = \sqrt{m^2 +  \lambda_{hv} v^2/2}$. 
In what follows, we will take an  agnostic attitude  to  the origin of the various terms
and will simply  scan over $\lambda_{hv}$ and $m_X$ to see their 
phenomenological implications (see also \cite{Kanemura:2010sh}). 
We set $\lambda =0$ for simplicity and allow for $\lambda_{hv}$
up to order one, which is consistent with perturbativity up to 10 TeV or so. 
Since the relevant rates  depend on $\lambda_{hv}^2 $, 
in the plots we will only display the absolute value of  $\lambda_{hv} $, whereas
it is understood that its contribution to $m_X$ can be of both signs.
Finally, reliable computations can be made as long as the 
unitarity constraint (\ref{uni}) is satisfied. For light $X$, we require the theory to be valid
up to the EW scale $\sim v$, whereas for heavy $X$ we take the cutoff to be of order $2 m_X$.

In our numerical analysis,
 we have adapted the code Micromegas \cite{Micromegas}  to our model in order to  
compute the thermal WIMP relic
abundance. We have also modified the direct detection cross section to take into account recent 
lattice results on the quark  structure of the nucleon, which appear more reliable than those
extracted from the pion--nucleon cross section. 
Concerning the Higgs physics, we use  the code Hdecay \cite{Hdecay}  
to compute  the Higgs decay widths.

\subsection{Direct detection and relic abundance constraints}

\begin{figure}
   \begin{center}
   \includegraphics[width=3.3in]{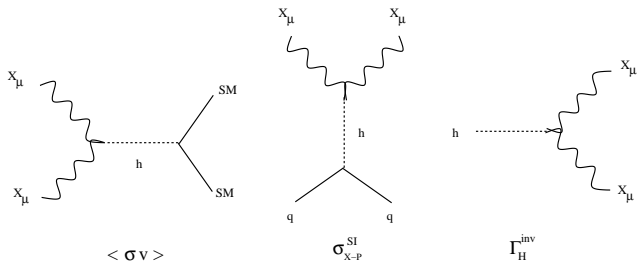}
   
          \caption{{\footnotesize
Feynman diagrams for DM  annihilation  (left), direct detection  (center) and invisible
 Higgs decay (right).
}}
\label{Fig:Feynman}
\end{center}
\end{figure}

\noindent
The Higgs portal coupling allows the vectorial dark matter to annihilate into the light Standard Model
species as well as to scatter off nuclei through the Higgs exchange. 
Fig.\ref{Fig:Feynman} displays  the relevant types of  Feynman diagrams.
(Numerically, we also take into account annihilation through the quartic vertex $X_\mu X^\mu h h $
which can be efficient at $m_X \gtrsim  m_h$, although we always find it  subdominant.)

Some aspects of the analysis  are  similar to those of the   scalar singlet extension of the SM.
The latter was studied in 
\cite{Silveira:1985rk,UM-P-91-54,McDonald:1993ex,McDonald:2001vt,Burgess:2000yq,Meissner:2006zh,
Davoudiasl:2004be,He:2008qm,Kanemura:2010sh,
Aoki:2009pf,Guo:2010hq,Barger:2010mc, Espinosa:2007qk,SungCheon:2007nw,Barger:2007im,Clark:2009dc,arXiv:1106.3097}.
Using this  framework,
some authors have  attempted  to explain the DAMA and/or COGENT 
excess \cite{Andreas:2008xy,Andreas:2010dz,Tytgat:2010bt}. Others have studied its implications for
indirect searches \cite{Yaguna:2008hd,Goudelis:2009zz,Arina:2010rb} and constraints from the
earlier XENON data \cite{Cai:2011kb,Biswas:2011td,Farina:2011bh}.
In what follows, we analyze the same constraints in the framework of vectorial dark matter.
We note that some phenomenological aspects of the vector Higgs portal  were 
studied in \cite{Kanemura:2010sh}, while vector dark matter in somewhat
different contexts was
considered in \cite{Hambye:2008bq},\cite{Hisano:2010yh}.

\begin{figure}
    \begin{center}
    \hspace{-1.cm}
   \includegraphics[width=3.in]{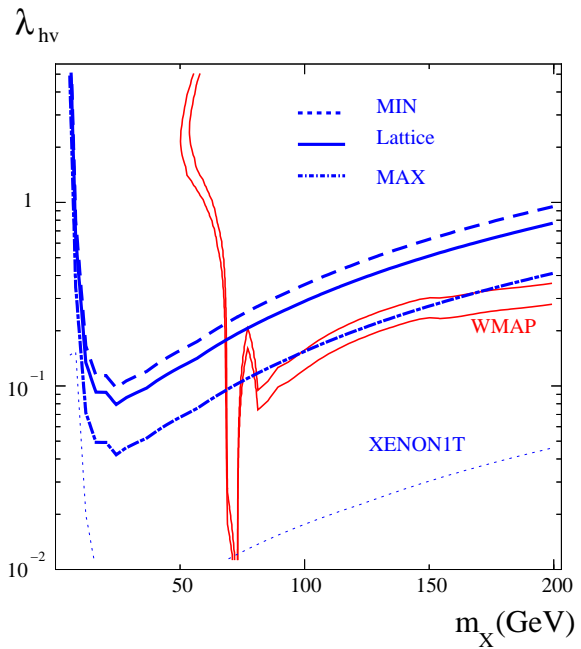}
          \caption{{\footnotesize
Parameter space allowed by WMAP (between the ``gull--shaped'' curves)  and  
XENON100    for $m_h = 150$ GeV, and prospects for XENON1T. 
}}
\label{Fig:Scan}
\end{center}
\end{figure}

The relic abundance of DM is dictated by  the $s$--channel  annihilation through the Higgs
exchange. For example, the  cross section for annihilation into fermions is given by
\begin{equation}
\langle  \sigma_{f \bar f} v   \rangle = { \lambda_{hv}^2 m_f^2 \over 48 \pi }~
{(1-m_f^2/m_X^2)^{3/2}  \over  (4 m_X^2 - m_h^2)^2  }  \;. 
\end{equation}
The calculational procedure is well known (see e.g. \cite{Kanemura:2010sh}) 
and here
we will only quote the result. The 5$\sigma$  WMAP--allowed \cite{WMAP}
parameter space \{$\lambda_{hv}, m_X$\}
for $m_h = 150$ GeV is shown in Fig.~\ref{Fig:Scan} between the two ``gull--shaped''
curves.
As in any Higgs portal model, DM  annihilation becomes much more efficient around the 
Higgs pole, $m_h \simeq 2 m_X$, which implies lower values of the coupling $\lambda_{hv}$. 
In Fig.~\ref{Fig:Scan},
we  also display the XENON100   constraint on the DM--nucleon interaction 
\cite{Aprile:2011ts}. 
The spin--independent cross section is given by
\be
\sigma^{SI}_{V-N} = \frac{\lambda_{hv}^2}{16 \pi m_h^4} \frac{m_N^4  f_N^2}{ (m_X + m_N)^2} \;,
\label{Eq:SigmaSI}
\ee
where $m_N$ is the nucleon mass and $f_N$ parametrizes the Higgs--nucleon coupling.
The latter subsumes  contributions of the light quarks $(f_L)$ and heavy quarks $(f_H)$,
$f_N=\sum f_L + 3 \times \frac{2}{27} f_H$. There exist  different estimations of this
factor and in what follows we will use the lattice result $f_N=0.326$
\cite{Young:2009zb} as well as the MILC results \cite{Toussaint:2009pz} which provide 
the minimal value $f_N=0.260$ and the maximal value $f_N=0.629$.
Since the cross section is rather sensitive to $f_N$, we dispay 3 curves corresponding
to these benchmark values. The region above these curves  is excluded by XENON100, so
the maximal allowed value for $\lambda_{hv}$  at $m_h = 150$ GeV  is  a few times $10^{-1}$.
This is still consistent with WMAP, especially around  the resonant annihilation region.
We also see that  essentially the entire parameter space will be probed by XENON1T.
Our results agree well with those of \cite{Kanemura:2010sh} (up to the XENON100 bound
which was not available at the time of publication of \cite{Kanemura:2010sh}).

Finally, we have checked that the unitarity constraint (\ref{uni}) is satisfied in the entire
WMAP--allowed band, apart from the region $\lambda_{hv} > 1$ where DM 
is light.
 For $\lambda_{hv} \sim 10^{-2}$, the cutoff is almost 2 orders 
of magnitude above $m_X$.

In what follows, we will use the XENON100 constraint based on the lattice 
evaluation of $f_N$.

\subsection{Implications  for the  Higgs search}

\begin{figure}
    \begin{center}
   \includegraphics[width=3.5in]{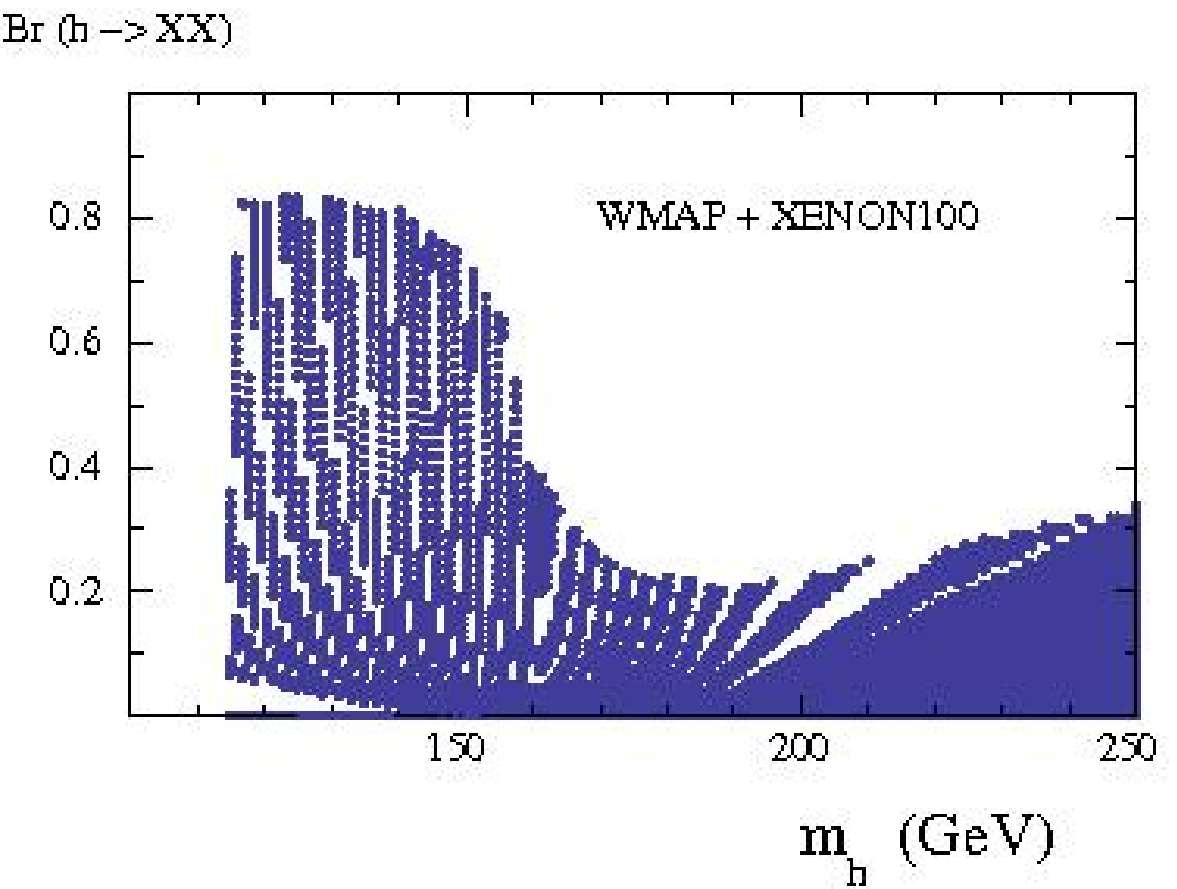}
   
      \includegraphics[width=3.5in]{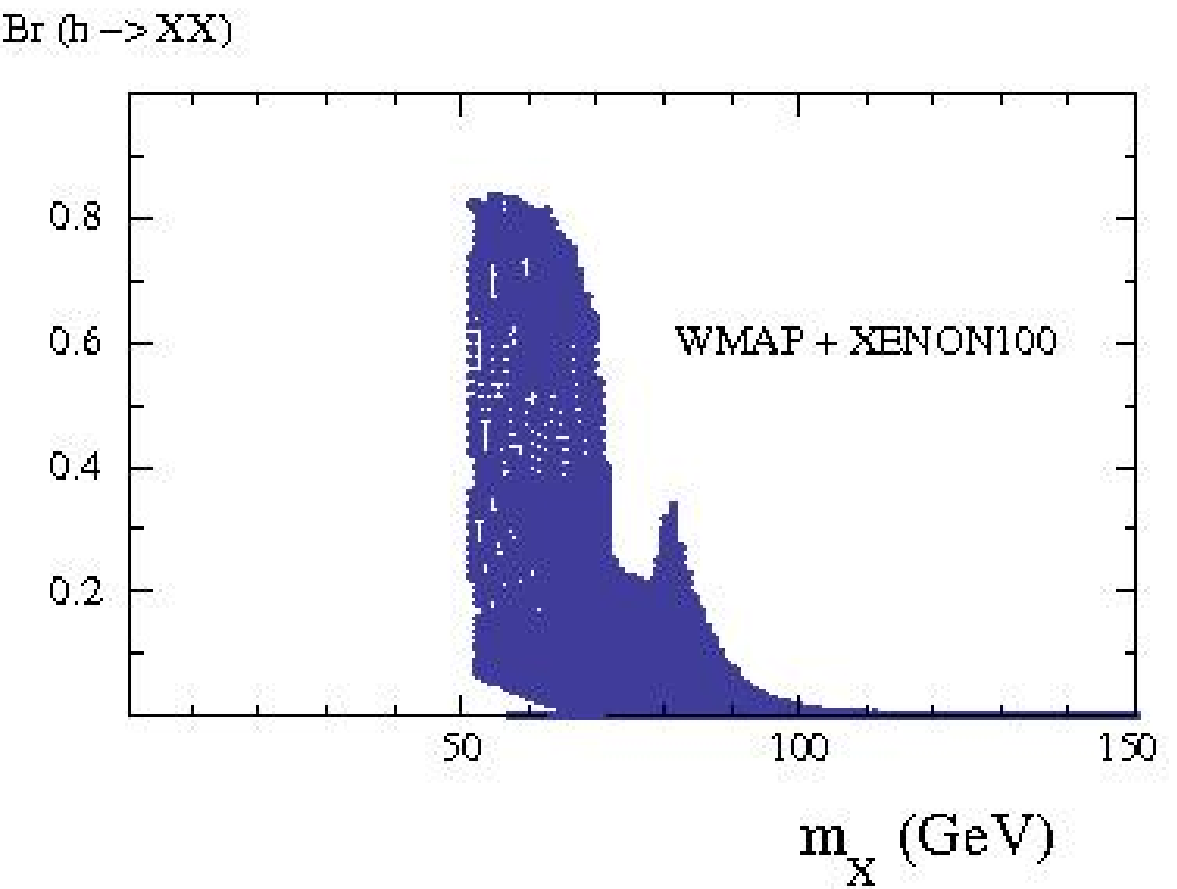}
   
          \caption{{\footnotesize
Branching ratio for the Higgs invisible decay 
as a function of the Higgs mass (top) and the dark matter mass (bottom).
The scan is over ($m_h$, $m_X$, $\lambda_{hv}$) and subject to the 
WMAP and XENON100 constraints.  }}
\label{Fig:BrInv}
\end{center}
\end{figure}

If the mass of the dark matter particle  is less than a half of the Higgs boson mass, 
dark matter can efficiently be produced in the ``invisible''
Higgs decay $h \rightarrow X X$.
This would affect the   Higgs decay branching ratios, while leaving the production
cross section unchanged. 
The partial  width for the Higgs decay into vector dark matter can be read off from (\ref{hXX}),
\begin{eqnarray}
\Gamma^{\rm inv}_{h \rightarrow X_{\mu} X_{\mu}} &=& \frac{\lambda^2_{hv} v^2 m_h^3}{256 \pi   m_{X}^4}
\left(
1-4 \frac{m_X^2}{m_h^2}+12\frac{m_X^4}{m_h^4}
\right) \nonumber\\
  &\times&  \sqrt{1-4 \frac{m_X^2}{m_h^2}} \;,
\label{Eq:Invwidth}
\end{eqnarray}
where $v = 174$ GeV.
We observe that this width grows fast with the Higgs mass, $\propto m_h^3$. This is because the decay
rate   into the longitudinal components of the vector field  or, equivalently, the would--be Goldstone
bosons, grows with energy due to the  derivative couplings of the latter. 
This is to be contrasted with the decay rate
into the usual  scalars which decreases with the Higgs mass.   
Thus even a small  coupling $\lambda_{hv}$ can lead to a large invisible branching fraction
for a light $X$.
In Fig.~\ref{Fig:BrInv}, we show the result of our scan over  ($m_h$, $m_X$, $\lambda_{hv}$)
taking into account the  WMAP and XENON100  constraints.
We find that  points with a large, up to 85\%, invisible branching ratio survive
for low Higgs masses.
There is a sharp fall in BR${}^{\rm inv}$ around 160 GeV, when the $h\rightarrow WW$
mode is in full force. The lightest $m_X$ consistent with 
WMAP and XENON100 is close to $m_h/2$,
in which case  $\lambda_{hv}$ is smaller than the gauge coupling and the invisible decay 
is subdominant. Nevertheless, BR${}^{\rm inv}$ can still be as large as 20-30\%.

\subsection{LHC constraints}

The invisible Higgs decay into dark matter can have a significant impact on the Higgs search
at the LHC. Since  the standard  Higgs decay modes
such as $\gamma \gamma$, $b \overline{b}$, etc. are not affected, 
the effective branching ratio for a given SM decay 
$ h \rightarrow  \alpha  \overline{\alpha}  $
decreases by a factor of ${\cal R} $, 
\be
{\rm BR}(h \rightarrow  \alpha  \overline{\alpha})_{\rm SM + X_{\mu}} = 
{\cal R}~ {\rm Br} (h \rightarrow \alpha \overline{\alpha})_{\mathrm{SM}} \;,
\ee
where
\be
{\cal R}= \frac{\Gamma^{\rm tot}_{\rm SM}}{\Gamma^{\rm tot}_{\rm SM}+ 
\Gamma (h \rightarrow X_{\mu} X_{\mu})} \;.
\ee
Since the production modes $qq/gg \rightarrow h$ remain the same, 
the ``measured'' cross section for the process  
$qq/gg \rightarrow h \rightarrow \alpha \overline{\alpha}$
is reduced by a factor of ${\cal R} $.
Consequently,  the event rates used in the  ATLAS/CMS analysis 
\cite{ATLAS,COMBI}
would be overestimated by  1/${\cal R}$
and the actual exclusion limits would be weaker. For example,  
$m_h \gtrsim  141$  GeV is 
no longer ruled out  by ATLAS/CMS.

\begin{figure}
    \begin{center}
   \includegraphics[width=3.5in]{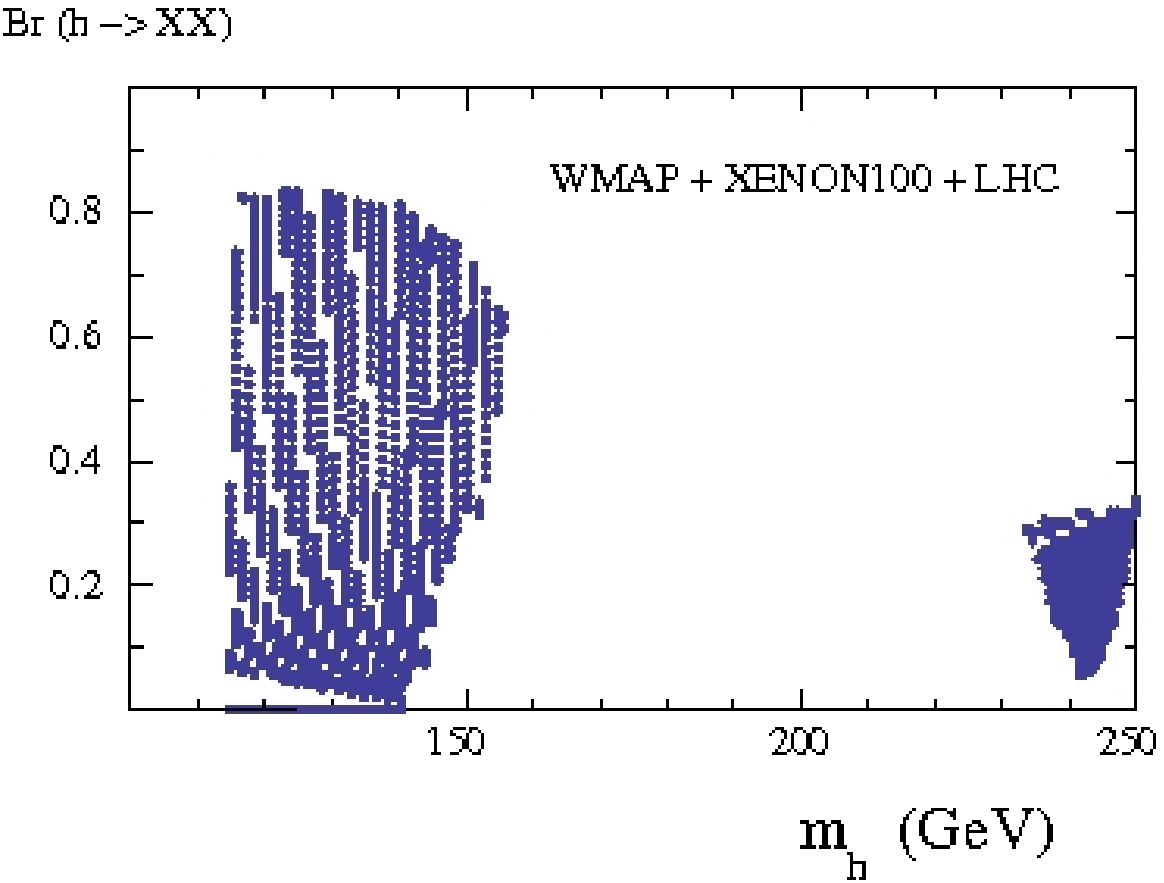}
   
      \includegraphics[width=3.5in]{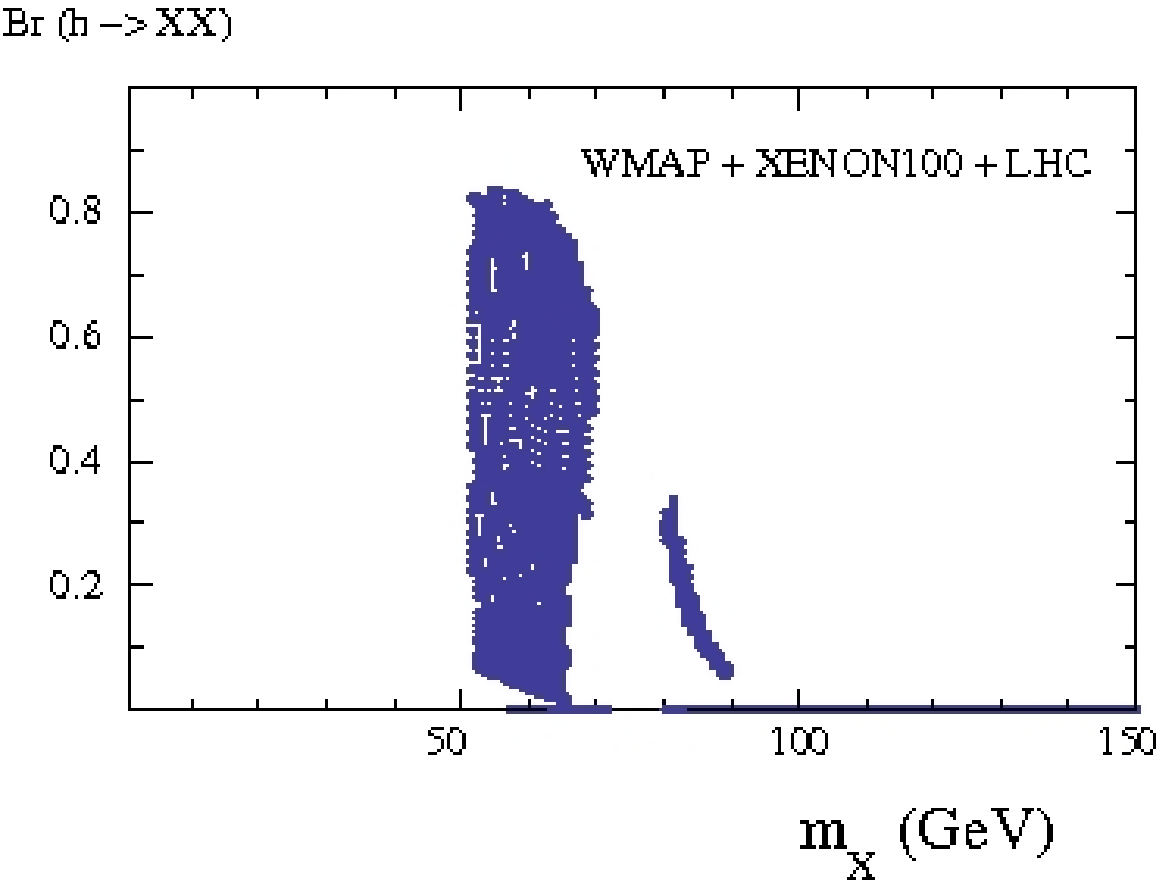}
   
          \caption{{\footnotesize
Branching ratio for the invisible Higgs decay  
as a function of the Higgs mass (top) and the dark matter mass (bottom).
The scan is over ($m_h$, $m_X$, $\lambda_{hv}$) and subject to the 
WMAP, XENON100 and ATLAS/CMS constraints. }}
\label{Fig:BrInvHExcl}
\end{center}
\end{figure}

\noindent
Fig.~\ref{Fig:BrInvHExcl} displays the result of our scan over  $m_X$ and $\lambda_{hv}$
subject to the WMAP, XENON100 and ATLAS/CMS constraints.
For each $m_h$, we find appropriate $m_X$ and $\lambda_{hv}$ such that 
the WMAP and XENON100 constraints are satisfied and  recalculate
the cross section for  $qq/gg \rightarrow h \rightarrow \alpha \overline{\alpha}$.
We find that  the invisible decay branching ratio can be as large as 85\% for $m_h$ up
to 150 GeV. For $m_h >160$ GeV it drops to 20-30\%, in which case the decay is almost SM--like
and the corresponding Higgs  mass range is largely excluded by  ATLAS/CMS.
Note, however, that the window $230 - 250$ GeV is still open and 
consistent with all the constraints.

The lower bound on  $m_X$ is determined roughly by $m_X \sim m_h/2$ for 
the minimal possible $m_h$, which is about 50 GeV.
Higher masses  $m_X >  m_h/2$, up to TeV,  are allowed by all the constraints. However,  the invisible
decay is inefficient in this case.

Finally, let us remark that if one gives up the assumption
of thermally produced vector dark matter, ${\rm BR}^{\rm inv}$  can be 
even larger.

\subsection{Comparison with scalar dark matter}

Analogous analyses have recently appeared in the framework of scalar dark matter 
interacting with the Standard Model through the Higgs portal  
\cite{Mambrini:2011ik,Hinv,Gonderinger:2009jp,Darkon,
 Weihs:2011wp}.
Although in many respects the scalar and vector cases are  similar, there is an 
important difference in the invisible decay of the Higgs.
While the DM scattering cross section on nucleons is the same in both cases
(for the same couplings and masses),
\be
\sigma^{SI}_{S-N} = \frac{\lambda_{hs}^2}{16 \pi m_h^4} \frac{m_N^4  f_N^2}{ (m_S + m_N)^2} \;,
\label{Eq:SigmaSI-S}
\ee
the invisible Higgs width for scalars is 
\be
\Gamma^{{\rm inv}}_{h\rightarrow SS}= \frac{\lambda_{hs}^2 v^2}{64 \pi  m_h} \sqrt{1 - 4 
{m_S^2 \over m_h^2}} \;,
\label{Eq:HiggsWidth-S}
\ee
where $\lambda_{hs}$ is the analog of $\lambda_{hv}$  (see Eq.~\ref{scalar-portal}). 
Unlike in the vector case  (\ref{Eq:Invwidth}), the decay rate decreases with the Higgs mass. 
As we mentioned before, this difference has to do with the nature of massive vector fields,
which absorb the would--be Goldstone bosons. This results in a higher decay rate at 
higher energies, i.e. for a heavier Higgs. The decay rate into vectors  is enhanced by 
 $m_h^4/m_X^4$, which is a large factor for a light $X$. In the region of interest to us,
$m_h \sim 2 m_X$ and the enhancement is an order of magnitude. There are also other 
numerical differences in the scalar versus vector dark matter. For example,
 WMAP requires 
roughly twice as large $\lambda_{hv}$ as $\lambda_{hs}$.
This is because the annihilation cross section for scalars is
\begin{equation}
\langle  \sigma_{f \bar f} v   \rangle = { \lambda_{hs}^2 m_f^2 \over 16 \pi }~
{(1-m_f^2/m_S^2)^{3/2}  \over  (4 m_S^2 - m_h^2)^2  }  \;, 
\end{equation}
which is triple that for vectors (due to averaging over polarizations)\footnote{
At low energies, the vector field can also be viewed as a set of 3 types of scalars which can only annihilate with scalars of the same type. This reduces the annihilation cross section
by a factor of 3.}
with   the same couplings and masses. 
Finally, the phase space coefficients in $\Gamma^{\rm inv}$
are different for vectors and scalars. 
We find that these factors compensate each other and the original
esimate $\sim m_h^4/m_X^4$ gives approximately the right answer. 
To give an example, at  $m_h=170$ GeV the WMAP/XENON100 constraints are satisfied for 
$\lambda_{hs}=0.13$ and $m_S=71$ GeV. The corresponding ${\rm BR}^{\rm inv}$ is 2\%.
The vector counterpart of this example is  $\lambda_{hv}=0.21$ and $m_X=72$ GeV,
which gives ${\rm BR}^{\rm inv}= 25$\%. As expected, the invisible decay is an order
of magnitude more efficient for the vectors. 

\begin{figure}
    \begin{center}
   \includegraphics[width=3.5in]{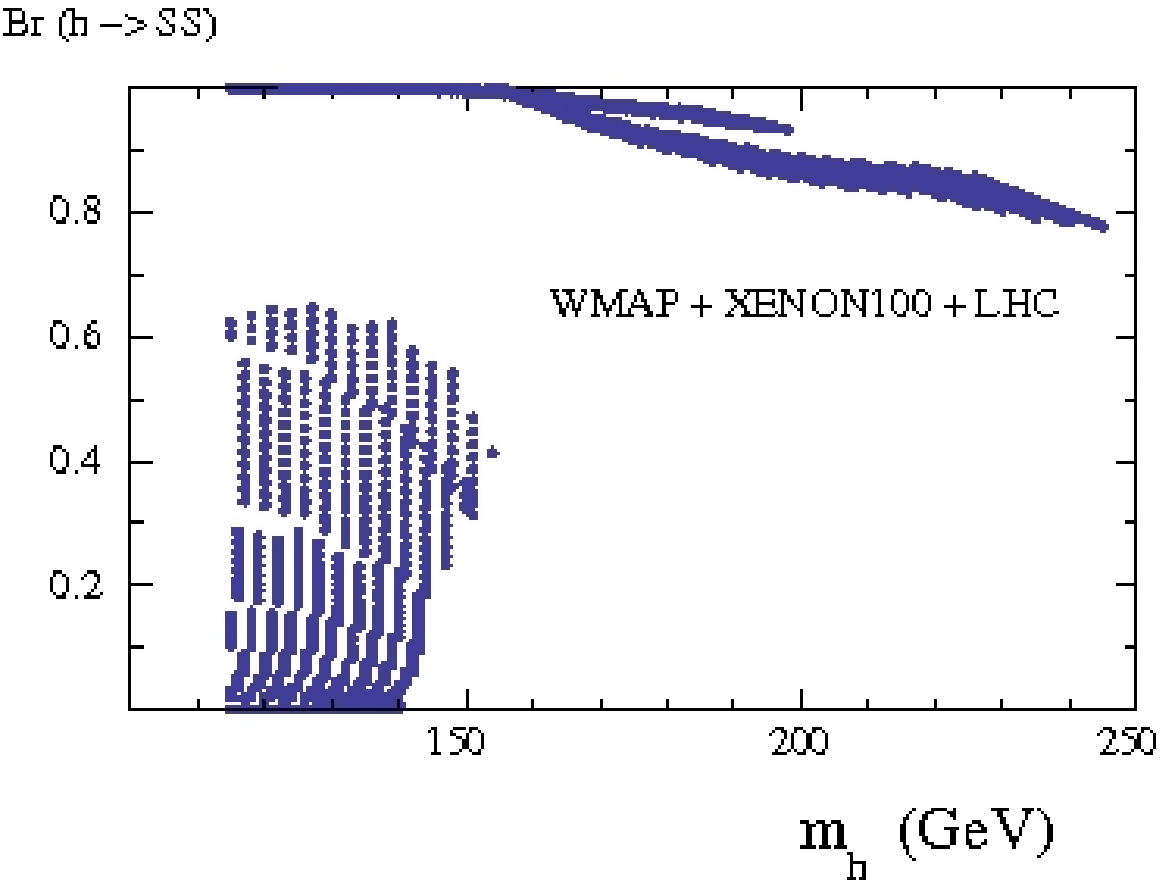}
   
      \includegraphics[width=3.5in]{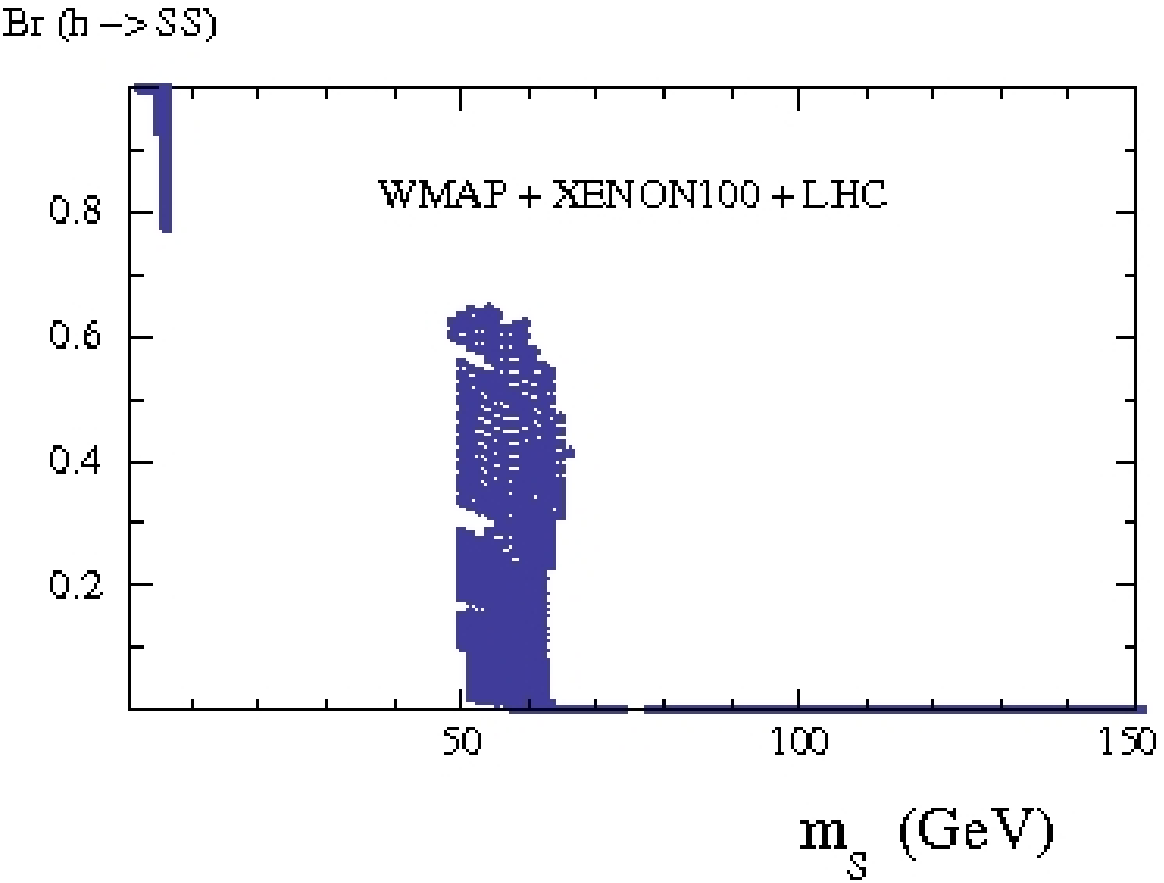}
   
          \caption{{\footnotesize
 Branching ratio for the invisible Higgs decay into scalars  
as a function of the Higgs mass (top) and the dark matter mass (bottom).
The scan is over ($m_h$, $m_S$, $\lambda_{hs}$) and subject to the 
WMAP, XENON100 and ATLAS/CMS constraints.    }}
\label{Fig:BrInvScalHExcl}
\end{center}
\end{figure}

Although $\Gamma^{\rm inv}$ for vectors is much larger than that for scalars, the 
effect on ${\rm BR}^{\rm inv} =\Gamma^{\rm inv} /( \Gamma^{\rm inv} + \Gamma^{\rm vis} ) $ 
is less significant if the Higgs decays predominantly into dark matter. 
At $m_h < 150$ GeV, the maximal ${\rm BR}^{\rm inv}$ for vectors is 85\% while that
for scalars is 65\% (excluding the special case of very light $\sim$ 5 GeV DM). 
We illustrate this in Fig.~\ref{Fig:BrInvScalHExcl} which
displays ${\rm BR}^{\rm inv}$ consistent with WMAP, XENON100
and ATLAS/CMS constraints for the scalar case.\footnote{The strip at the top corresponds
to very light scalar DM studied in \cite{Andreas:2010dz}. } 
Note that, up to the above reservations, 
the Higgs mass window $230 - 250$ GeV is not available since the 
invisible decay is inefficient for a heavy Higgs. 

An important difference between the vector and scalar cases is that 
in the former case very light dark matter is not allowed by unitarity.
For $m_X \sim 5$ GeV and $\lambda_{hv} \sim 1$, the unitarity cutoff 
is less 10 GeV meaning that the Higgs portal is not a sensible description
of the complete model. Additional states must be invoked to restore unitarity,
which in turn would modify our DM analysis. This also applies to the case of 
$m_X \sim 50$ GeV, $\lambda_{hv} \sim 2$ WMAP--allowed region 
(Fig.~\ref{Fig:Scan}), where the cutoff is below the EW scale.

\section{Conclusion}

We have studied the possibility that dark matter is of vectorial nature and
it communicates with the Standard Model through the Higgs portal. 
We find this possibility well motivated because abelian gauge theories with the
minimal field content, which is  necessary to  render the vector 
fields massive, 
possess a   natural $Z_2$ parity (charge conjugation)
making dark matter stable.  
Vector dark matter is consistent with cosmological and direct detection 
constraints, and can lead to invisible Higgs decay. The latter is 
enhanced     compared to the scalar dark matter case and
provides an efficient way to hide the Higgs at the LHC.
A combination of  sensitive  direct detection experiments (XENON1T)
and further collider searches will probe most of parameter space of 
this scenario.

\noindent
{\bf Acknowledgements. } 
The authors would like  to thank particularly M. Kado and J.B. Devivie
for their valuable help with the combined ATLAS/CMS analysis.
Y.M. thanks  E. Bragina and   the Magic Monday Journal Club 
 for very useful discussions and comments.
We are also grateful to S. Pukhov for his help in solving
technical problems related to micrOMEGAs. 
This  work was
supported by the French ANR TAPDMS {\bf ANR-09-JCJC-0146} 
and the Spanish MICINNÕs Consolider-Ingenio 2010 Programme 
under grant  Multi- Dark {\bf CSD2009-00064}.
HML is partially supported by the CERN-Korean fellowship.



\end{document}